\documentclass[aps,prc,preprint]{revtex4-1}

\usepackage{graphics,epsfig}
\usepackage{rotating}

\begin{document}


\title{Separable potential model for meson-baryon interaction beyond the S-wave}

\author{Vojt\v{e}ch Krej\v{c}i\v{r}\'{i}k}
\email{vkrejcir@umd.edu}

\affiliation{Maryland Center for Fundamental Physics, Department of Physics, \\
 University of Maryland, College Park, MD 20742-4111}

\begin{abstract}
A model for low-energy meson-baryon interaction in the strange sector is presented.
The interaction is described in terms of separable potentials with multiple partial waves considered.
A general solution of Lippmann-Schwinger equation for the scattering of spin zero and spin one-half particles is derived.

Next, the general framework is applied to the $\overline{K}N$ sector in a simple model with  only the S- and P-waves taken into account.
The separable potential is designed to match the chiral perturbation theory
at lowest nontrivial order.
It is shown that although a simple model with three free parameters works well for the S-wave,
it fails to reproduce the P-wave features of kaon-nucleon physics.
Most importantly, the P-wave interaction is too weak to
express a resonant behavior that could be identified as $\Sigma(1385)$ resonance.

\end{abstract}


\maketitle

\section{Introduction}

The description of low-energy meson-baryon interaction in the strange sector is a highly puzzling problem.
The direct application of the effective theory approach \cite{Weinberg}, which was  successful in the pion-nucleon sector,
i.e., baryonic chiral perturbation theory (for review see \cite{BernardKaiserMeissner}),
is problematic.
The key physical issue is the presence
of the $\Lambda(1405)$ resonance below the $\overline{K}$-$N$ threshold \cite{PDG}.
The existence of a resonance implies the need to work to all orders in perturbation theory
and therefore procedures alternative to standard $\chi$PT are required.

A possible way to proceed is  via the multi-channel Lippmann-Schwinger equation
with the interaction described by separable potentials \cite{KaiserSiegelWeise, CieplySmejkal07, CieplySmejkal10}.
The physics based on chiral symmetry of QCD is reflected in the design of the respective separable potentials.
They are designed to match the amplitudes obtained in chiral perturbation theory up to given order $O(p^n)$.
In the hypothetical world of  very low-quark masses, amplitudes obtained by iterating the Lippmann-Schwinger equation with such potentials
are equal to the amplitudes derived in $\chi$PT up to a given order in the chiral expansion $O(p^n)$.
Note that the $\chi$PT is constructed as an effective theory of QCD in the regime of low momenta and low quark masses.
However, in the physical world with a relatively high $s$ quark mass, the connection to the fundamental theory of
strong interactions---the quantum chromodymanics---is more subtle.

The suggested approach has, on the other hand, the advantage that the Lippmann-Schwinger
equation is exactly solvable; the originally complicated system of coupled integral equations simplifies
to an algebraic equation and a set of integrals.
Additionally, some features of nuclear medium, for example, Pauli blocking \cite{WaasKaiserWeise} and/or self-energy effects \cite{Lutz}
can be straightforwardly incorporated into the separable potential model.
Kaon-nucleon amplitudes enriched with in-medium effects may then be used to determine an effective in-medium
kaon-nuclear potential  \cite{CieplyFriedman.., CieplySmejkal12}, as well as to study other low-energy processes involving
kaon-nuclear interaction, for example, the hypernuclear production \cite{KrejcirikCieplyGal, CieplyFriedmanGalKrejcirik}.

In their recent work, Ciepl\'y and Smejkal \cite{CieplySmejkal10, CieplySmejkal12}
were able to fit a large set of low-energy experimental data (threshold branching ratios,
kaonic hydrogen shift and width, cross sections to various channels for kaon incident momentum up to 200 MeV)
with a very simple model combining the chiral dynamics
with separable potential approach considering only the $L=0$ partial wave.
They were also able to analyze the properties of $\Lambda(1405)$ resonance.
However, the model of Ref. \cite{CieplySmejkal10} has an important limitation---it considers only the S-wave contribution.
Although this turned out to be sufficient for the kaon incident momentum up to 200 MeV,
the inclusion of higher partial waves becomes necessary if one wants to go to higher kaon momenta.
Moreover, the P-wave interaction is expected to play an important role in the formation of deeply bound
$K^-$-nuclear states \cite{WeiseHartle, GazdaMares}.
Although the authors of \cite{GazdaMares} used the more fundamental model of Ref. \cite{CieplySmejkal12} for the S-wave interaction,
they relied on purely phenomenological parametrization for P-wave interaction. Thus, the improvement of the understanding
of the P-wave part of meson-baryon interaction in the strange sector is certainly desirable.

The goal of this paper is to extend the separable potential model of Ref. \cite{CieplySmejkal10} for meson-baryon interactions
in the strange sector to include the effects of the P-wave.
With the P-wave part included in the separable potential, one should
gain access
to phenomena which are  inaccessible if only S-wave is included, for example,
the angular distribution of the cross section.
Moreover, one might expect that the P-wave resonance $\Sigma(1385)$ could be dynamically generated using the P-wave potential in
a similar way as $\Lambda(1405)$, as was studied by Ciepl\'y and Smejkal.

The paper is organized as follows. The general formalism of spin zero spin one-half particle scattering is summarized
in Sec. 2. In Sec. 3, the general form for a multi-channel separable potential model is discussed and the solution of
the Lippmann-Schwinger equation is derived for such potentials. In Sec. 4, the potential for kaon-proton
scattering is constructed to match the $\chi$PT up to lowest nontrivial order. Fits to the available experimental data and
discussion of obtained results follows in the Sec. 5.

\section{Formalism of spin zero spin one-half scattering}

In this section, the formalism describing two-particle scattering---one with spin one-half
and one with spin zero---is summarized.
To be concrete, only the interactions that are both time-reversal and parity invariant are considered;
all formulas  are given in the center-of-mass frame of reference.

The most general form of the scattering amplitude  is a $2 \times 2$ spin matrix \cite{Goldberger}:
\begin{equation}
{\bf f}({\bf p} \rightarrow {\bf p'}) = \tilde{f}({\bf p} \rightarrow {\bf p'})
+ i  \, {\bf \sigma} \cdot \hat{\bf p} \times \hat{\bf p}' \,\, \tilde{g}({\bf p} \rightarrow {\bf p'}) \,.
\label{amplgeneral}
\end{equation}
In (\ref{amplgeneral}), ${\bf p}$ and ${\bf p'}$ stand for the initial and final CMS momentum. A hat above a vector indicates a unit vector in the
respective direction.
The  spin non-flip and spin flip amplitudes are denoted as $\tilde{f}$ and $\tilde{g}$.

Due to parity invariance, the total angular momentum $J$ and orbital angular momentum $L$ are separately conserved
during the scattering. Thus, for a given initial orbital momentum $L$ (a given partial wave), there are two independent S-matrix
elements---one for $J=L+1/2$, and one for $J=L-1/2$---that fully characterize the scattering process.
Therefore, one can write down the generalized partial wave expansion using the projection operators into the respective subspaces $J=L\pm 1/2$.
The amplitude reads:
\begin{equation}
{\bf f}({\bf p} \rightarrow {\bf p'}) =\sum\limits_{L} \left( 2L+1 \right) \left( f^{L+}({\bf p} \rightarrow {\bf p'}) \,  \Lambda^{L+} + f^{L-}({\bf p} \rightarrow {\bf p'})\,  \Lambda^{L-}  \right) P_L(\hat{\bf p} \cdot \hat{\bf p}') \,,
\label{amplpartial}
\end{equation}
where $\Lambda^{L \pm}$ is a projection operator to a subspace of total angular momentum $J=L\pm 1/2$:
\begin{eqnarray}
\Lambda^{L+} &=& \frac{1}{2L+1} \left( L+1 + {\bf \sigma} \cdot {\bf L}   \right) \,, \nonumber\\
\Lambda^{L-} &=& \frac{1}{2L+1} \left( L - {\bf \sigma} \cdot {\bf L}   \right) \,. \nonumber
\end{eqnarray}

Simple algebraic manipulations allow  one to find a partial wave expansion for spin flip and
spin non-flip amplitudes ($\tilde{f}$, $\tilde{g}$) in terms of
amplitude projections for given total and orbital angular momenta ($f^{L+}$, $f^{L-}$):
\begin{eqnarray}
\tilde{f} &=&  \sum\limits_{L}  \left[ (L+1) \, f^{L+} + L \, f^{L-}  \right] P_L(\hat{\bf p} \cdot \hat{\bf p}') \,, \label{amplSpinNonflip}\\
\tilde{g} &=& \sum\limits_{L} \left[ f^{L+} - f^{L-}  \right] P_L'(\hat{\bf p} \cdot \hat{\bf p}') \,. \label{amplSpinFlip}
\end{eqnarray}

Note, that if the spin-orbit coupling  is zero, amplitude $f^{L+}$ is equal to $f^{L-}$.
In this case, the partial wave expansion for spin non-flip amplitude $\tilde{f}$ coincides with the standard
one for the scattering of two spinless
particles and the spin flip amplitude $\tilde{g}$ vanishes.

In this parametrization, the differential cross section for unpolarized beam and target reads:
\begin{equation}
\frac{{\rm d}\sigma}{{\rm d}\Omega} ({\bf p} \rightarrow {\bf p'})
= |\tilde{f}({\bf p} \rightarrow {\bf p'})|^2 + \sin^2(\theta) \,\, |\tilde{g}({\bf p} \rightarrow {\bf p'})|^2 .
\label{difcsection}
\end{equation}

If only $L=0$ and $L=1$ are included---the model which will be investigated later in the paper---the
differential and total cross sections  read:
\begin{eqnarray}
\frac{{\rm d}\sigma}{{\rm d}\Omega}  &=&    \left( \left|f^{0+}\right|^2 + \left|2f^{1+}+f^{1-}\right|^2 \cos^2\theta + \left|f^{1+}-f^{1-}\right|^2 \sin^2\theta \right. + \nonumber\\
&& \left. \left( f^{0+}(2f^{1+} + f^{1-})^* +  (f^{0+})^*(2f^{1+} + f^{1-})  \right) \cos\theta \right), \label{sigmaforL01}\\
\sigma^{\rm tot} &=& 2\pi \left( 2 \left|f^{0+}\right|^2 + \frac{2}{3} \left|2f^{1+}+f^{1-}\right|^2 + \frac{4}{3} \left|f^{1+}-f^{1-}\right|^2   \right).
\end{eqnarray}

The angular distribution of the differential cross section is often parameterized in powers of $\cos \theta$:
\begin{eqnarray}
\frac{{\rm d}\sigma}{{\rm d}\Omega} &\approx&  A_0 + A_1 \cos{\theta} +A_2 \cos^2{\theta} + \dots  \,, \\
A_0 &=& \left|f^{0+}\right|^2 + \left|f^{1+}-f^{1-}\right|^2 \,, \nonumber \\
A_1 &=&   2\, {\rm Re} \left[ f^{0+}(2f^{1+} + f^{1-}) \right]  \,,  \nonumber \\
A_2 &=& \left|2f^{1+}+f^{1-}\right|^2 -  \left|f^{1+}-f^{1-}\right|^2  \,. \label{assymetries_formula}
\end{eqnarray}
Note that $A_0$, $A_1$, and $A_2$ are the only three coefficients that are non-zero if only S- and P- waves are considered.

\section{Separable potentials and a solution of \\ Lippmann-Schwinger equation}

In this section, a general form of a multi-channel separable potential between spin
one-half and spin zero particles is discussed.
It is shown that, even in this relatively complicated case, the solution of a Lippmann-Schwinger equation simplifies to
an algebraic problem.
To cover the most general situation when particle types can change during the process,
the multi-channel approach to the problem is employed from the beginning;
individual channels are labeled $(ai)$, where $a$ stands
for the type of spin one-half particle (baryon), and $i$ stands for the spin zero particle (meson).

Knowing the general form of the scattering amplitude (\ref{amplpartial}), the potential will be taken  to have a form:
\begin{equation}
V_{(ai)\rightarrow(bj)}({\bf p} \rightarrow {\bf p'}) = \sum\limits_{L} (2L+1) \left( V_{(ai)\rightarrow(bj)}^{L+} \Lambda^{L+} + V_{(ai)\rightarrow(bj)}^{L-} \Lambda^{L-}  \right) P_L(\hat{\bf p} \cdot \hat{\bf p}') \, g_{(ai)}^L(p) \, g_{(bj)}^L(p') \,,
\label{SepPotential}
\end{equation}
where $g_{(ai)}^L(p)$ is a form factor corresponding to the channel $(ai)$ and the partial wave $L$.

The Lippmann-Schwinger equation reads:
\begin{eqnarray}
T_{(ai)\rightarrow(bj)}({\bf p} \rightarrow {\bf p'}) &=& V_{(ai)\rightarrow(bj)}({\bf p} \rightarrow {\bf p'})  \nonumber \\
& +&\sum\limits_{(ck)} 2 \mu_{(ck)}  \int {\rm d}^3q \frac{V_{(ai)\rightarrow(ck)}({\bf p} \rightarrow {\bf q})\, T_{(ck)\rightarrow(bj)}({\bf q} \rightarrow {\bf p'})}{p_{(ck)}^2-q^2+ i \epsilon} \,,
\label{LSequationgeneral}
\end{eqnarray}
where the sum is over all possible intermediate channels $(ck)$, $p_{(ck)}$ is the on-shell momentum in the intermediate channel
and $\mu_{(ck)}$ is
its reduced energy.

Having the potential in the separable form (\ref{SepPotential}), it is natural to use the following ansatz for the T-matrix:
\begin{equation}
T_{(ai)\rightarrow(bj)}({\bf p} \rightarrow {\bf p'}) = \sum\limits_{L} (2L+1) \left( T_{(ai)\rightarrow(bj)}^{L+} \Lambda^{L+} + T_{(ai)\rightarrow(bj)}^{L-} \Lambda^{L-}  \right) P_L(\hat{\bf p} \cdot \hat{\bf p}') \,g_{(ai)}^L(p) \, g_{(bj)}^L(p') \,.
\label{SepTMatrix}
\end{equation}

In order to prove that the separable potential (\ref{SepPotential}) and the T-matrix ansatz (\ref{SepTMatrix}) actually
transform the Lippmann-Scwhinger equation into an algebraic equation, the crucial
thing is to show that the integral in  Eq. (\ref{LSequationgeneral})
actually splits into two pieces, one for $J=L+1/2$ and one for $J=L-1/2$, and preserves the separability for each piece.
The separation of angular and radial part of the integral gives:
\begin{eqnarray}
\sum\limits_{L L'} \sum\limits_{(ck)} 2 \mu_{(ck)}   \int {\rm d}q & q^2 \frac{(2L+1)(2L'+1)}{p_{(ck)}^2-q^2+i \epsilon} \, g_{(ai)}^L(p) \, g_{(ck)}^L(q) \, g_{(ck)}^{L'}(q) \,  g_{(bj)}^{L'}(p')  \nonumber \\
\int {\rm d}\Omega_{\bf q} & \left( V_{(ai)\rightarrow(ck)}^{L+} \Lambda^{L+} + V_{(ai)\rightarrow(ck)}^{L-} \Lambda^{L-}  \right) P_L(\hat{\bf p} \cdot \hat{\bf q}) \nonumber \\
  &  \left( T_{(ck)\rightarrow(bj)}^{L'+} \Lambda^{L'+} + T_{(ck)\rightarrow(bj)}^{L'-} \Lambda^{L'-}  \right) P_{L'}(\hat{\bf q} \cdot \hat{\bf p}') \,. \nonumber
\end{eqnarray}
The orthogonality of Legendre polynomials
and the properties of projection operators
are key ingredients in the proof of separability.
Finally, one gets two independent sets of matrix equations in the channel space, one for $J=L+ 1/2$ and one for $J=L-1/2$:
\begin{equation}
T_{(ai)\rightarrow(bj)}^{L \pm} = V_{(ai)\rightarrow(bj)}^{L\pm} + V_{(ai)\rightarrow(ck)}^{L\pm} G_{(ck)}^L  T_{(ck)\rightarrow(bj)}^{L\pm} \,.
\label{LSequationLpm}
\end{equation}
$G^L$ is a diagonal matrix with elements given by the integral:
\begin{equation}
G_{(ck)}^L  = 4 \pi \mu_{(ck)}  \, \int {\rm d}q \, q^2 \frac{ \left(g_{(ck)}^L(q)\right)^2 }{p_{(ck)}^2-q^2+i\epsilon} \,.
\end{equation}

The relation between the T-matrix elements $T_{(ai)\rightarrow(bj)}^{L\pm}$
and the scattering amplitudes  $f_{(ai)\rightarrow(bj)}^{\pm}$ is
\begin{equation}
f_{(ai)\rightarrow(bj)}  = -4\pi \, \sqrt{\mu_{(ai)}\mu_{(bj)}} \,\, T_{(ai)\rightarrow(bj)}\,. \nonumber
\end{equation}
From here one straightforwardly obtains all the information about the scattering process using formulas derived in the previous section.

In the case with only S- and P-waves considered, the potential is of the form:
\begin{eqnarray}
V_{(ai)\rightarrow(bj)} &=& V^{0+} g^0_{(ai)} g^0_{(bj)} +3 \left( V^{1+} \Lambda^{1+} +V^{1-} \Lambda^{1-} \right) g^1_{(ai)} g^1_{(bj)}  \cos{\theta} \nonumber\\
 &=&  V^{0+} g^0_{(ai)} g^0_{(bj)} + \nonumber\\
 && \left[\left( 2 V^{1+} + V^{1-} \right) \cos{\theta} + \left( V^{1+} - V^{1-} \right) i {\bf \sigma} \cdot \hat{\bf p} \times \hat{\bf p}' \right]  g^1_{(ai)} g^1_{(bj)} \,.
\label{SepPotentialL01}
\end{eqnarray}
Analogous expansions hold for T-matrix and scattering amplitude as well.
Overall, there are three independent equations of the form (\ref{LSequationLpm}) for $T^{0+}$, $T^{1+}$, and $T^{1-}$.

\section{The construction of separable potentials for kaon-proton scattering}

The separable potential (\ref{SepPotentialL01}) used in the calculation will be constructed to match the chiral perturbation theory up to order
$O(p)$, the lowest nontrivial order, in this section. Thus, a brief review of relevant parts of baryon $\chi$PT is in order.

The first order  chiral Lagrangian \cite{OllerMeissner} reads:
\begin{eqnarray}
\mathcal{L}^{(1)} &=& i \left<  \overline{B}  \gamma_{\mu} \left[ D^{\mu} , B \right]   \right> - M_0 \left<  \overline{B} B \right>  \nonumber\\
  && - \frac{D}{2} \left<  \overline{B}  \gamma_{\mu}\gamma_{5} \left\{ u^{\mu} , B \right\}   \right> - \frac{F}{2} \left<  \overline{B}  \gamma_{\mu}\gamma_{5} \left[ u^{\mu} , B \right]   \right>  \,.
\end{eqnarray}

\begin{figure}
\begin{center}

\includegraphics[width=2in]{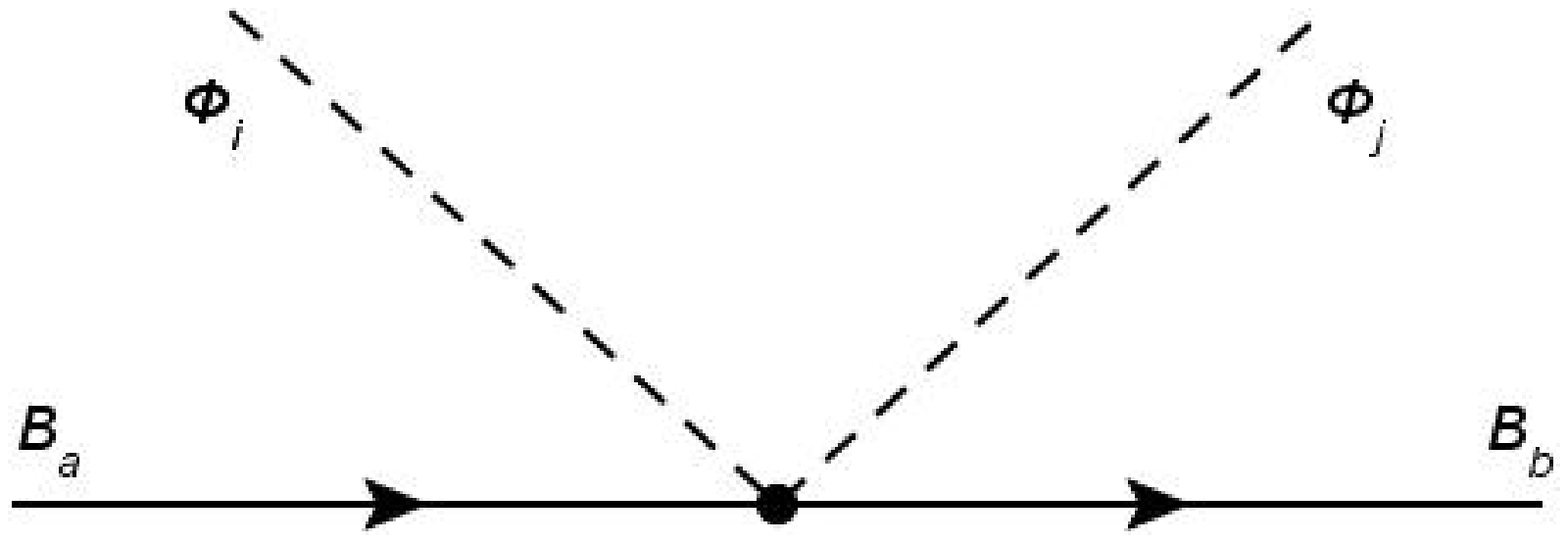}\\
\includegraphics[width=2in]{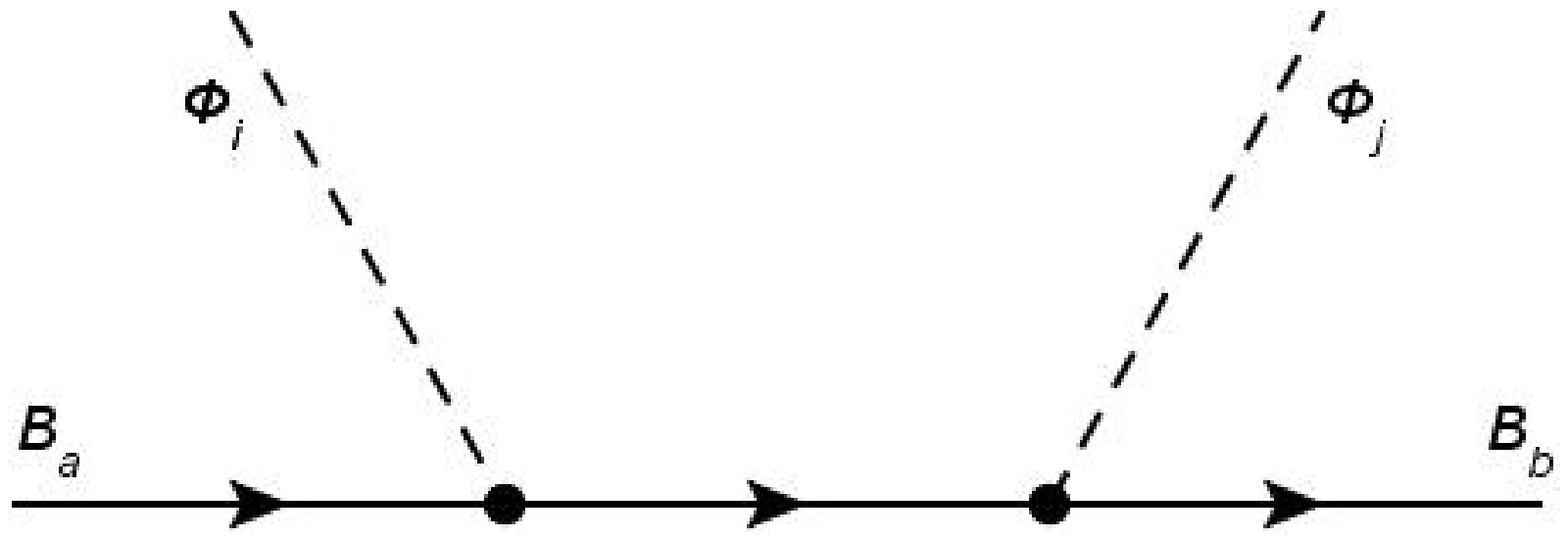}
\includegraphics[width=2in]{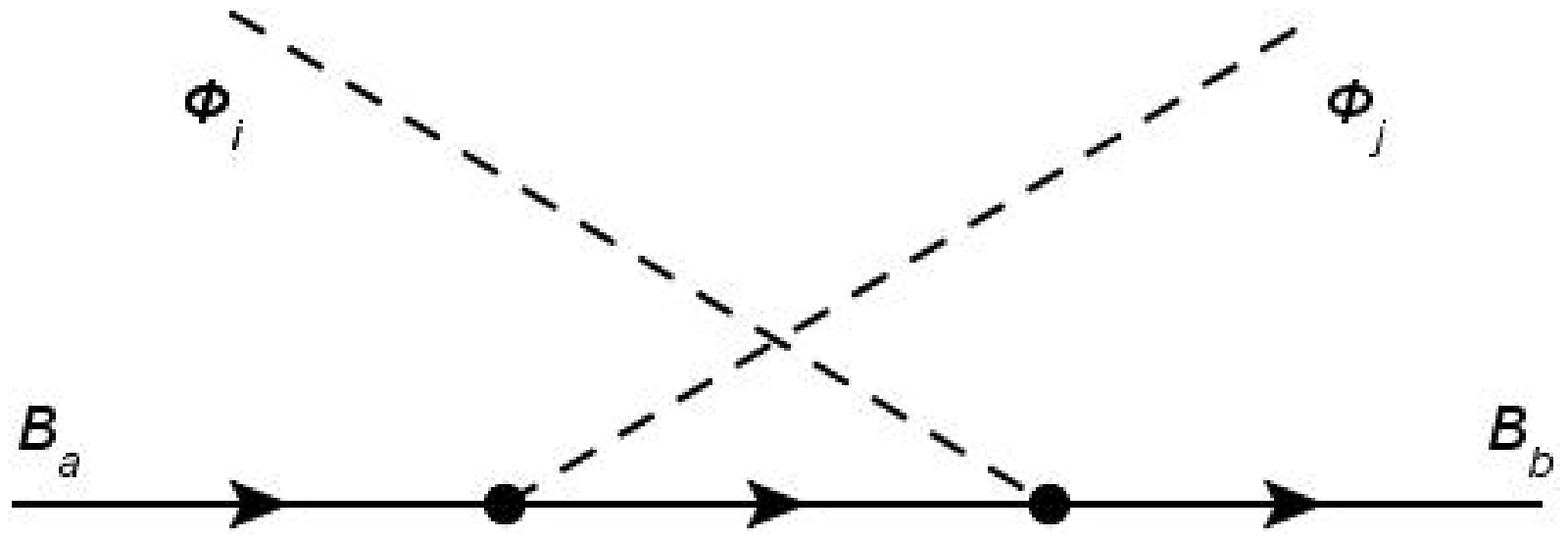}

\caption{Feynman diagrams representing the Weinberg-Tomozawa, direct (s-channel), and crossed (u-channel)
contributions to the meson-baryon scattering.}
\label{diagrams}

\end{center}
\end{figure}

For meson-baryon scattering, one gets three principally different contributions from diagrams summarized in Fig. \ref{diagrams}.
The first one represents the contact interaction coming from $O(p^1)$ Lagrangian, the so-called Weinberg-Tomozawa term (WT);
it is a leading contribution to the S-wave amplitudes.
The other two diagrams correspond to the direct (s-channel) and crossed (u-channel) processes
built from vertexes from $O(p^1)$ Lagrangian.
At first sight, these seem to be of order $O(p^2)$. However, the non-relativistic baryon propagator $1/k^0$ is of order $O(p^{-1})$ making the
leading behavior of the diagram to be $O(p^1)$. These diagrams represent the leading order contribution to the P-wave amplitudes.
Some of the amplitudes may be found in the literature \cite{OllerMeissner, BorasoyNisslerWeise,  BrunsMaiMeissner, MaiMeissner}, and all of them can be
reconstructed from the potentials that will be presented below.

The separable potentials (\ref{SepPotentialL01}) are constructed to match the chiral perturbation theory:
\begin{equation}
V_{(ai)\rightarrow(bj)} = \frac{1}{4(2\pi)^3} \sqrt{\frac{1}{s \mu_{(ai)} \mu_{(bj)}}} \,\, \mathcal{M}_{(ai)\rightarrow(bj)},
\label{VtoM}
\end{equation}
where $\mathcal{M}_{(ai)\rightarrow(bj)}$ are Lorenz invariant scattering amplitudes obtained in $\chi$PT up to a given order $O(p^n)$;
only the leading order $O(p^1)$ is considered in this paper.

Since  the S- and P-waves are considered in the model, two form factors are needed; one for $L=0$
and one for $L=1$. For simplicity, Yamaguchi-type form factors \cite{Yamaguchi1} are used in the calculation:
\begin{eqnarray}
g_{(ai)}^0(p) &= & \frac{1}{1+\frac{p^2}{\alpha_{(ai)}^2 }} \,,\label{YamFF0} \\
g_{(ai)}^1(p) &= & \frac{p}{\left(1+\frac{p^2}{\alpha_{(ai)}^2} \right)^{3/2}} \,. \label{YamFF1}
\end{eqnarray}
$\alpha_{(ai)}$ characterize the the range of the interaction in the particular channel.

The use of (\ref{SepPotentialL01}), (\ref{VtoM}), (\ref{YamFF0}), and (\ref{YamFF1}) immediately leads to the potentials $V^{L\pm}$.
Recall that for the $L=1$ part, the one power of the momentum $p$ ($p'$) is already included in the form factor
$g^1_{(ai)}(p)$ (\ref{YamFF1}) and therefore does not appear in the respective potentials $V^{1p}$ (\ref{potV1p}), and $V^{1-}$(\ref{potV1m}).
The potentials up to order $O(p^1)$ read:
\begin{eqnarray}
V_{(ai)\rightarrow(bj)}^{0+} &= \mathcal{N} & \left[    -\frac{1}{8} \left( E_i + \frac{E_i^2-m_i^2}{2M_i^2} + E_j + \frac{E_j^2-m_j^2}{2M_j^2} \right) \,\,\,  {\bf \mathcal{C}}^{WT}_{(ai)\rightarrow(bj)}  \right]  \,, \label{potV0p}\\
V_{(ai)\rightarrow(bj)}^{1+} &= \mathcal{N} &   \left[    -\frac{1}{9}   \left( \frac{1}{m_i+m_j}  \right)  {\bf \mathcal{C}}^{crossed}_{(ai)\rightarrow(bj)}  \right] \,, \label{potV1p} \\
V_{(ai)\rightarrow(bj)}^{1-} &= \mathcal{N} &   \left[ \frac{1}{6} \left( \frac{1}{m_i+m_j}   \right) \,\,\,  {\bf \mathcal{C}}^{direct}_{(ai)\rightarrow(bj)}   + \frac{1}{18}   \left( \frac{1}{m_i+m_j} \right) \,\,\,  {\bf \mathcal{C}}^{crossed}_{(ai)\rightarrow(bj)}  \right] \,. \label{potV1m}
\end{eqnarray}
The constant $\mathcal{N}$ guaranteeing proper relativistic flux normalization reads:
\begin{equation}
\mathcal{N} = \frac{1}{2f_{\pi}^2} \frac{1}{(2\pi)^2} \sqrt{\frac{M_a M_b}{s \mu_{(ai)} \mu_{(bj)} }} \,.
\end{equation}
Matrices $\mathcal{C}^{\dots}$ are summarized in the appendix. They determine how the channels are coupled between each other.

A brief comment about the $V^{1\pm}$ potentials is in order.
In the leading order amplitude in $\chi$PT, there is a sum of meson energies $E_i+E_j$ in the denominator.
However, it leads to a possible unphysical divergence in the deep subthreshold region.
This divergence is trackable back to the form of the fully relativistic baryon propagator, which diverges for $k_{\mu}k^{\mu}=-M^2$.
The complication here is due to the fact that the calculation is restricted
solely to the tree diagrams at the level of $\chi$PT. This divergence, eventually, disappears if one
continues to higher orders.
In the presented model, the spurious divergence is avoided by replacing the meson energy $E_i$ by meson mass $m_i$ wherever it
appears in the denominator, as was suggested in Ref. \cite{CieplySmejkal12}.

The potential for the S-wave $V^{0+}$ part is identical to the leading part of the potential used by Ciepl\'y and Smejkal \cite{CieplySmejkal10} and
Kaiser et al. \cite{KaiserSiegelWeise}.
The P-wave contribution $V^{1+}$, $V^{1-}$ has not been considered by these authors.

\section{Fit to the low-energy $K^-$-$p$ data and the discussion of obtained results}

The comparison of the chirally motivated separable potential model from the previous section
to the experimental low-energy data is presented and discussed here.

The model developed in the previous section contains, in its full complexity, a substantial number of free parameters.
For example, there are in principle 20 different inverse range parameters $\alpha_{(ai)}^{0,1}$ characterizing form factors $g_{(ai)}^{0,1}(p)$ (\ref{YamFF0}), (\ref{YamFF1}).
More parameters would appear if one considers the second order chiral Lagrangian.
However, it is useful to keep the analysis simple and straightforward.
Thus, the model is restricted to the first order chiral Lagrangian and
the same inverse range parameter is used for all channels:
$\alpha^S$ for the S-wave form factors (\ref{YamFF0}), and $\alpha^P$ for the P-wave form factors (\ref{YamFF1}).
In the leading order $\chi$PT, the decay constant $f_{\pi}$ is the same for all mesons in the pseudoscalar octet (pions, kaons, $\eta$); its
value is not constrained experimentally and is thus subject to fit.
Overall, there are three free parameters to be specified: inverse ranges $\alpha^S$ and $\alpha^P$,
and $f_{\pi}$ controlling the strength of the interaction.

From the point of view of the partial wave analysis,
the low-energy experimental data
may be divided into three subcategories.
First, the threshold branching ratios \cite{ADMartin} and kaonic hydrogen properties \cite{BazziEtAl} are influenced entirely by the S-wave physics.
Cross sections are influenced by both S- and P-waves
(and, naturally, higher partial waves as energy of the collision increases), yet they are non-zero even if the P-wave is neglected.
And finally, properties of the angular distribution  of the differential cross sections require the presence of the P-wave.
All of them are considered in the analysis.
Since the $\Lambda(1520)$, which is a D-wave resonance,
emerges at the kaon incident momentum around 400 MeV, the validity of the model containing only S- and P-wave is certainly limited to below this value;
this paper considers only the scattering data up to 300 MeV \cite{Ciborowski, Evans, Sakitt, HumphreyRoss, Mast, Bangerter, Watson},
since there is both a significant effect of the P-wave while the D-wave is
expected to be negligible.
The key novelty of this paper is the focus on the effects of P-wave interaction, therefore the  attention is given to angular distribution
of the differential cross section.
The asymmetry in the angular distribution, which is parameterized by the quantity $A_1/A_0$  (\ref{assymetries_formula}),
begins to be observable at kaon incident momenta above 200 MeV \cite{Ciborowski, Evans, Mast, Bangerter}.
The experimental data on angular distributions are, unfortunately, very imprecise and
available only for channels $\pi^- \Sigma^+$,  $\pi^+ \Sigma^-$,  $K^- p$,  and $\overline{K}^0 n$.

In the fitting of free parameters,
 a simple minimization of $\chi^2$ is problematic;
 the $\chi^2$ fit weights most importantly the data points that are most precise. In this case, the threshold branching
ratios, whose uncertainties are orders of magnitude smaller than uncertainties of other data points,
would dominate the fit, whereas the properties of angular distribution, which depend dominantly on P-wave physics,
would be almost irrelevant because their uncertainties are big.
Since the main focus of this paper is the P-wave interaction, the following procedure,
which emphasizes the P-wave physics, is adopted.
First, the threshold characteristics and total cross section at low kaon momenta (at 100 MeV for $\pi^- \Sigma^+$, and  $\pi^+ \Sigma^-$,
at 110 MeV for $K^- p$, and $\overline{K}^0 n$, and at 120 MeV for $\pi^0 \Lambda$, and  $\pi^0 \Sigma^0$)
are fitted with only the S-wave interaction taken into account. It is justified by the fact that the P-wave interaction is
negligible for such low energies.
In this procedure, the free parameters controlling the S-wave interaction, $\alpha^S$ and $f_{\pi}$, are set.
Next, with the S-wave potential fixed,
the total cross sections at higher kaon momentum (300 MeV for all channels) and asymmetries in the angular distribution
($A_1/A_0$ for momenta 225, 250, 275, and 300 MeV for channels
$\pi^- \Sigma^+$,  $\pi^+ \Sigma^-$,  $K^- p$, and  $\overline{K}^0 n$)
are fitted with full potential in order to get the $\alpha^P$.

\begin{figure}
\begin{center}

\includegraphics[width=2.8in]{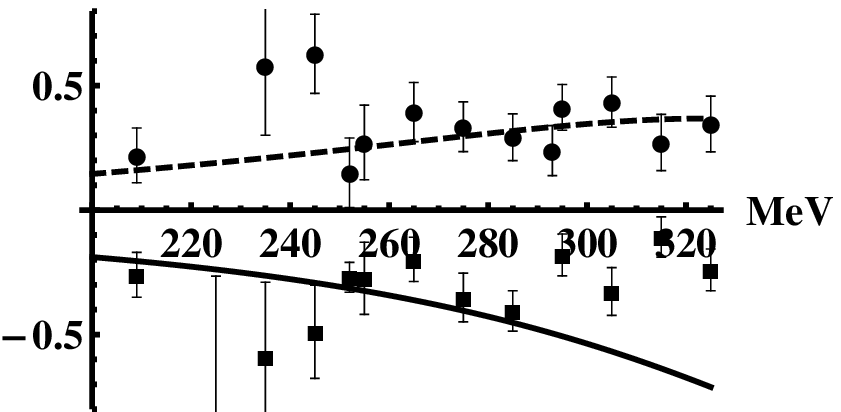}
\includegraphics[width=2.8in]{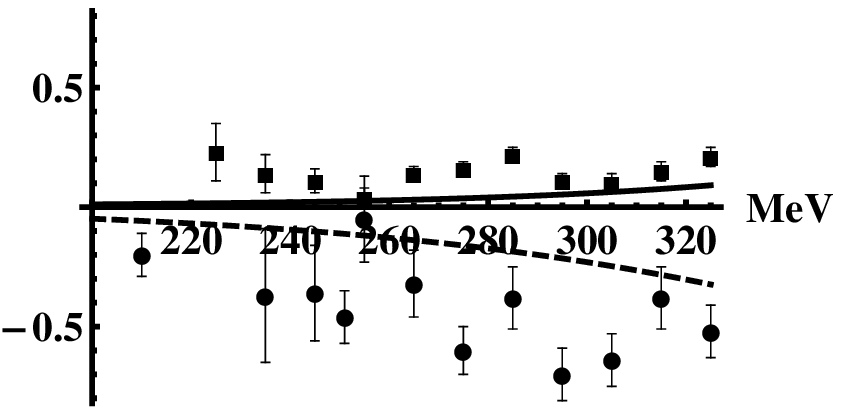}

\caption{Asymmetries $A_1/A_0$ in the differential cross sections. Channels $\pi^- \Sigma^+$ (full line, boxes),
 and  $\pi^+ \Sigma^-$ (dashed line, circles) are in the first graph; channels $K^- p$ (full line, boxes),
 and  $\overline{K}^0 n$ (dashed line, circles) are in the second graph.}
\label{asymetries}

\end{center}
\end{figure}

\begin{figure}
\begin{center}

\includegraphics[width=2.8in]{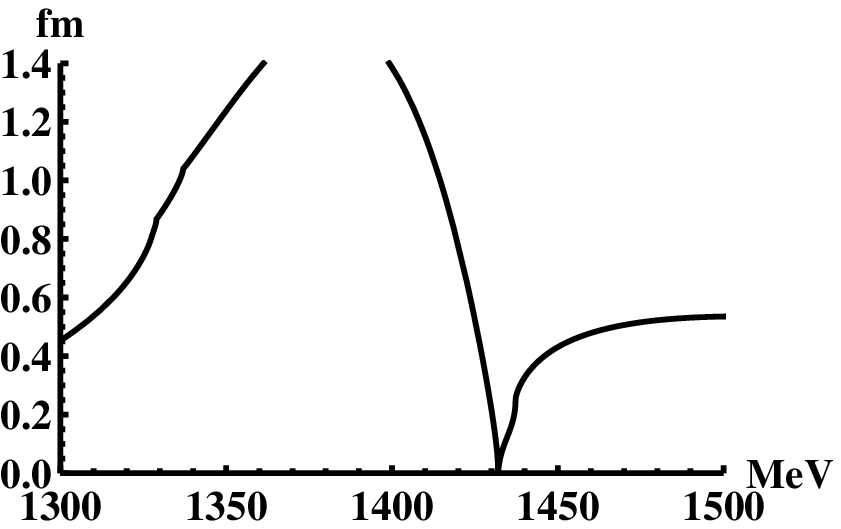}
\includegraphics[width=2.8in]{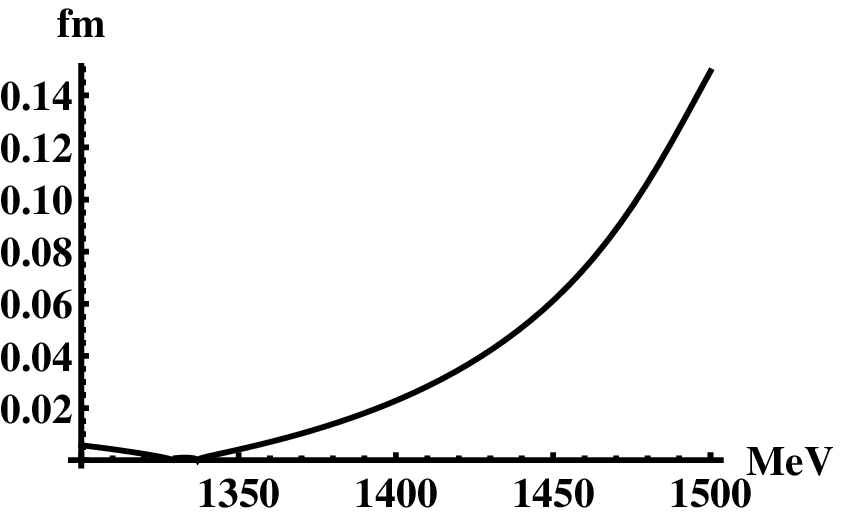}

\caption{Absolute values of isoscalar $\pi\Sigma$ amplitudes for $0^+$ (left) and isovector $\pi\Sigma$ amplitude for $1^+$ (right) partial waves.}
\label{resonances}

\end{center}
\end{figure}

Best results were obtained for the following values of free parameters:
$\alpha^S=736\, {\rm MeV}$, $f_{\pi}=116.6\, {\rm MeV}$, and $ \alpha^P=1353 \,{\rm MeV}$;
the overall $\chi^2/N =  4.3$.
The fact that the $\chi^2/N$ is well above 1 is not a problem  because the presented model does not aspire to be the complete
description of nature.
The comparison with experimental data is summarized in Table \ref{table_res2}, and Figs. \ref{asymetries}-\ref{resonances}.

\begin{table}[h!]
\caption{Kaon-nucleon threshold data.}
\label{table_res2}

\begin{tabular}{ c |ccc |c }
 && fit && exp. \cite{ADMartin, BazziEtAl} \\ \hline
 $\gamma$  && 2.36 && $2.36\pm 0.04$   \\ \hline
 $R_c$ && 0.637  &&$0.664 \pm 0.011$   \\ \hline
 $R_n$  && 0.178 &&$0.189\pm 0.015$   \\ \hline
 $\Delta E$  && $-296$ eV &&$-283 \pm 42$ eV  \\ \hline
 $\Gamma$ &&  $761$ eV &&  $541 \pm 111$ eV  \\
 \end{tabular}

\end{table}

\begin{figure}
\begin{center}

\includegraphics[width=2.8in]{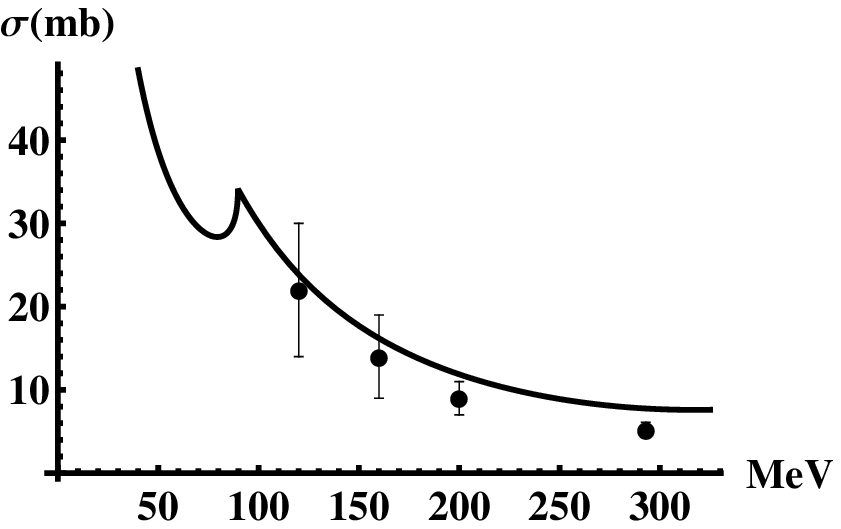}
\includegraphics[width=2.8in]{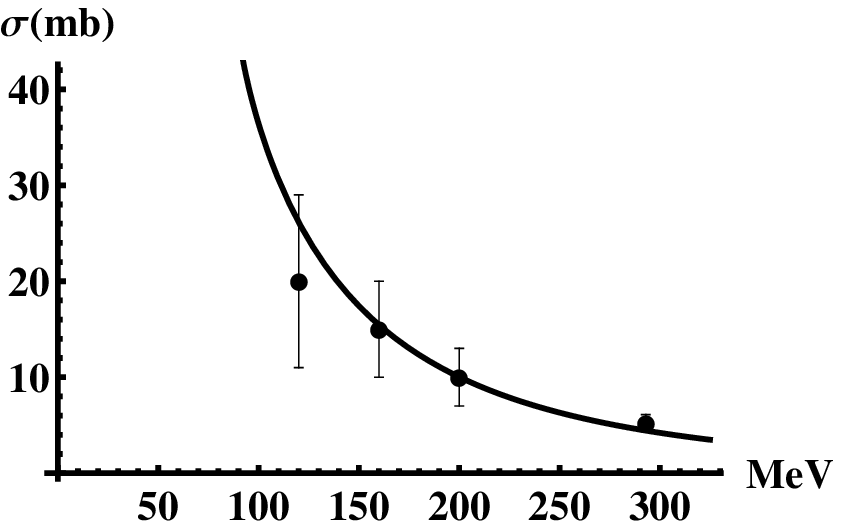}\\
\includegraphics[width=2.8in]{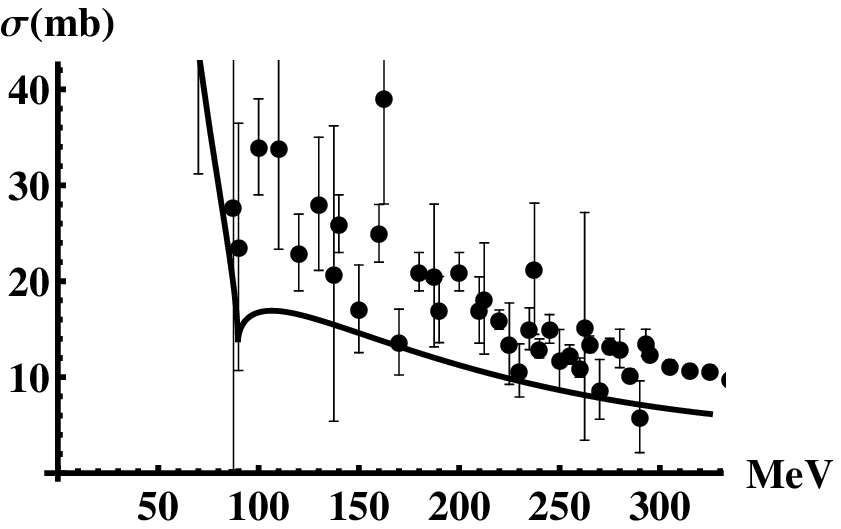}
\includegraphics[width=2.8in]{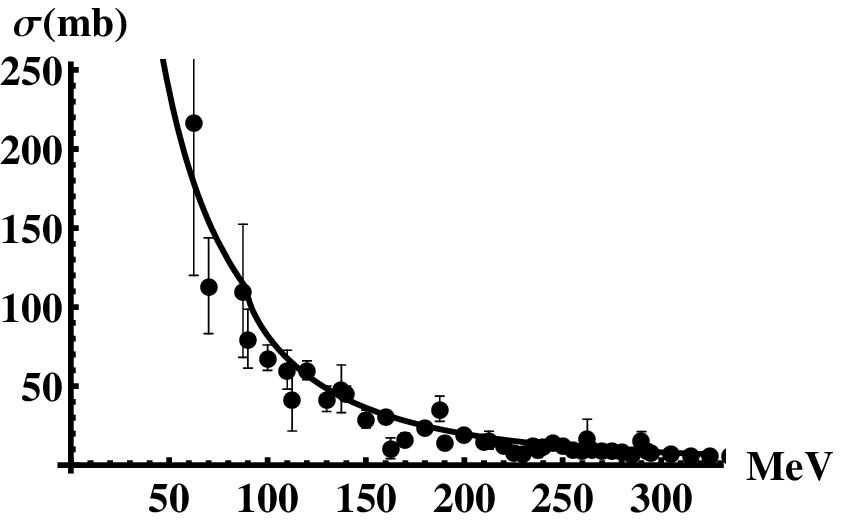}\\
\includegraphics[width=2.8in]{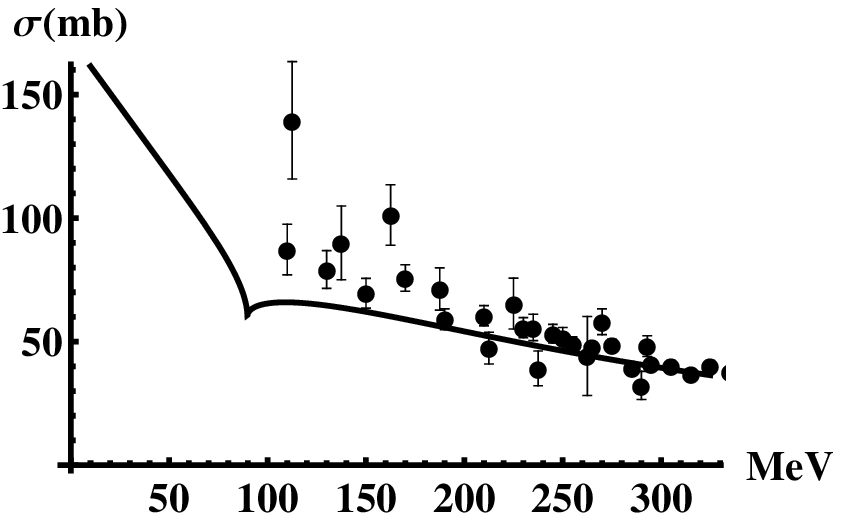}
\includegraphics[width=2.8in]{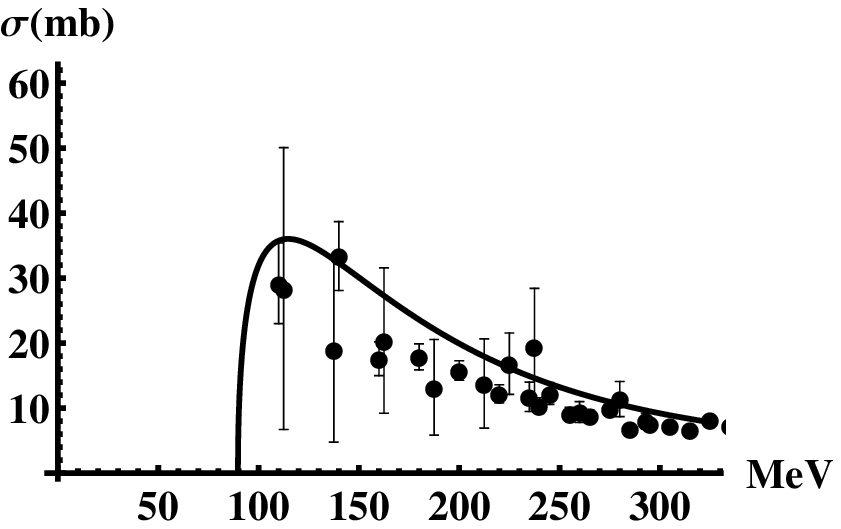}

\caption{The total cross section for channels (from top left) $\pi^0 \Lambda$, $\pi^0 \Sigma^0$, $\pi^- \Sigma^+$, $\pi^+ \Sigma^-$,
 $K^- p$, and $\overline{K}^0 n$. Experimental data are from \cite{Ciborowski, Evans, Sakitt, HumphreyRoss, Mast, Bangerter, Watson}. }
\label{Csections}

\end{center}
\end{figure}

As is seen in table \ref{table_res2} and figure \ref{Csections}, the agreement of the model with the data
is satisfactory for both the threshold data and total cross sections; the $\chi^2/N=2.9$ for the first part of the fit.
Note that it was expected since
it was already shown by Ciepl\'y and Smejkal \cite{CieplySmejkal10} that the chirally motivated separable
potential model considering only
the S-wave reproduces the wide range of low-energy experimental data.

However, the agreement is considerably worse in the P-wave sector.
In Fig. \ref{asymetries}, the asymmetries in the angular distribution of the differential
cross sections $A_1/A_0$ are shown for all four channels where data are available.
Although the experimental data are reproduced sufficiently well in the $\pi^- \Sigma^+$,
and  $\pi^+ \Sigma^-$ channels, the model fails for the $K^- p$, and  $\overline{K}^0 n$ channels.
Note that the sign of the asymmetry is correct, but the absolute value is too small.
It suggests that the P-wave potential is too weak.

The notion that the P-wave potential
motivated by the $O(p^1)$ chiral Lagrangian
 is too weak is enforced if one looks at the possible resonance  in the
$\pi\Sigma$ amplitudes (see Fig. \ref{resonances}).
The isoscalar amplitude in the $0^+$ partial wave clearly shows a resonant structure, which can be identified
with the $\Lambda(1405)$  (as was discussed more extensively in \cite{CieplySmejkal10}).
On the other hand, in the isovector $1^+$ partial wave, where the $\Sigma(1385)$ resonance lies, there is no
resonant behavior observed.

\begin{figure}
\begin{center}

\includegraphics[width=2.8in]{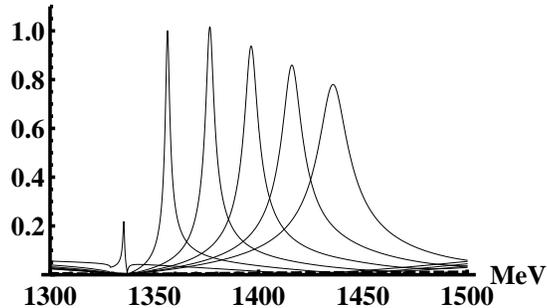}

\caption{Absolute values of isovector $\pi\Sigma$ amplitude for $1^+$ (right) partial waves obtained with
artificially strengthened P-wave interaction.
Curves correspond to the values of  $P_{scale}$ from $-3.9$ to $-4.4$ with peaks moving from left to right.
Dashed line corresponds to the $P_{scale}=1$.}
\label{resonances2}

\end{center}
\end{figure}

In order to check this hypothesis, the dependence of the $\pi\Sigma$ isovector amplitude
on the strength of the P-wave interaction was studied in a simplest possible way.
The P-wave potential (\ref{potV1p}), (\ref{potV1m}) was multiplied by the new  parameter $P_{scale}$, which controls
the strength of the P-wave interaction, while keeping the parameters $f_{\pi}$, $\alpha^S$, and $\alpha^P$  fixed
to the values obtained in the fit.
As is seen in Fig. \ref{resonances2},
the resonant structure above the $\pi\Sigma$ threshold appears for $P_{scale}^{RES}\approx-3.9$ and with increasing strength of the potential
moves towards higher energies.

\begin{figure}
\begin{center}

\includegraphics[width=2.8in]{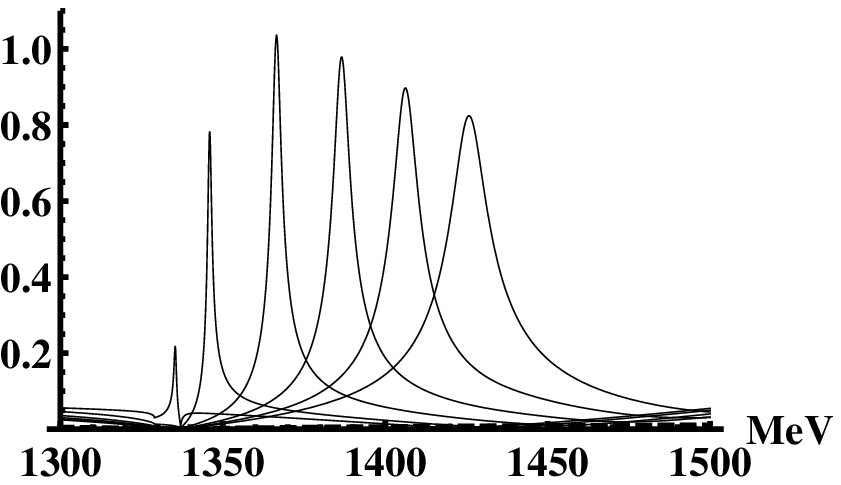}
\includegraphics[width=2.8in]{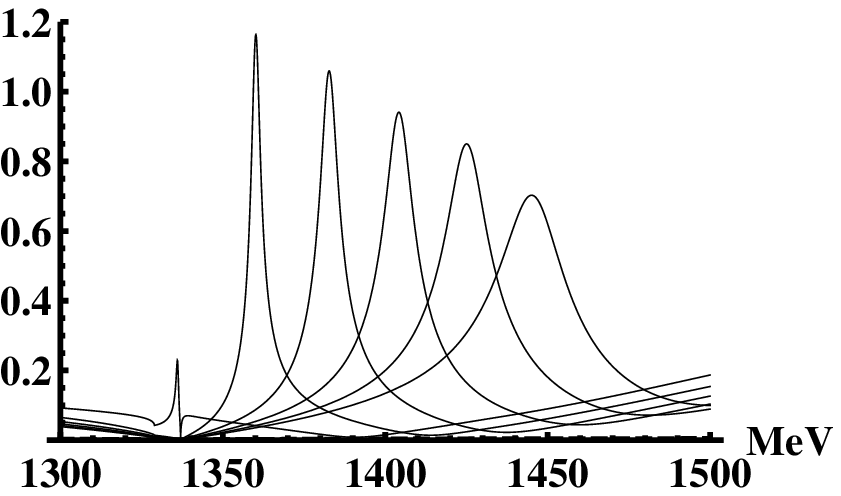}

\caption{Absolute values of isovector $\pi\Sigma$ amplitude for $1^+$ (right) partial waves obtained with
artificially strengthened P-wave interaction. Left figure corresponds to the $\alpha^P=1100 \,{\rm MeV}$ and values of
 $P_{scale}$ from -6.7 to -7.7. Right figure corresponds to the $\alpha^P=900 \,{\rm MeV}$ and values of
 $P_{scale}$ from -10.5 to -13.0.}
\label{resonances22}

\end{center}
\end{figure}

The numerical value of the $P_{scale}^{RES}$ where the resonant structure appears depends on the value of $\alpha^P$.
With the decreasing value of inverse range parameter $\alpha^P$, the absolute value of $P_{scale}^{RES}$ increases.
Graphs analogous to the one in figure \ref{resonances2} for $\alpha^P=1100 \,{\rm MeV}$ and $\alpha^P=900\, {\rm MeV}$
are presented in Fig. \ref{resonances22}; corresponding values of $P_{scale}^{RES}$ are $-6.7$ and $-10.5$, respectively.

Note  that the model of this paper is based only on the first order chiral Lagrangian.
In their latest version,
the model of Ciepl\'y and Smejkal \cite{CieplySmejkal12} was based on the second order
chiral Lagrangian and had seven free parameters to describe the S-wave physics only.
Similarly, a more sophisticated potential containing the P-wave interaction can be developed
if one considers second order Lagrangian and
allows different inverse range parameters for different channels.
Such a model could, in principle, be able to capture the desired resonant behavior in
the $1^+$ partial wave  if the respective low-energy constants coming from
the second order chiral Lagrangian were big enough.

The approach using a more sophisticated model, however, leads to a substantial increase in the number of free parameters.
Note that there is already a certain level of freedom in the formulation of the model itself.
For example, the functional form of the form factors (\ref{YamFF0}), (\ref{YamFF1}) is not constrained
by any underlying theory and the Yamaguchi-type form was chosen for simplicity.
The vast space of free parameters combined with poor quality of the data would make
any interpretation  of any result  obtained quite problematic.
On the other hand, better measurements of angular distribution asymmetries or other quantities
that would more firmly constrain the P-wave interaction would certainly
gain a more firm physical ground to the extended model.

The role of chiral perturbation theory is more subtle in the presented model.
Since the $\overline{K}N$ system is outside the regime of validity
of chiral perturbation theory---it was
the original motivation for an alternative approach at the first place---$\chi$PT serves only as a guideline
in the construction of the respective potentials.
Note, on the other hand, that the very use of $\chi$PT for the P-wave physics in the strange sector may raise
an important question.
It was established phenomenologically \cite{ADMartin, JKKim} that the $\Sigma(1385)$ couples strongly to the
$\pi\Lambda$ channel, weakly to the $\pi\Sigma$ channel, and negligibly to the $\overline{K}N$ channel.
It suggests that the $SU(3)$ flavor symmetry is badly broken and therefore the use of chiral physics
may be misguided---even if it serves only as a motivation.
Recall, however, that chirally motivated models were successful in the description of various S-wave phenomena in the strange
sector and thus the use of chiral physics as a guideline for in the construction of P-wave models seems to be a natural extension. 
From this point of view, the introduction of a new scaling parameter $P_{scale}$, which on one hand shifts the model away from $\chi$PT and
on the other hand improves the agreement with the experimental data, does not seem to be unreasonable.

In summary, it was shown that the model based on the solution of the Lippmann-Schwinger equation with the interaction
described by the separable potential (developed in Sec. 3)
is able to capture the physics of the P-wave interaction.
It was also shown that, although working
quite well for the S-wave, the model based  on the chiral Lagrangian at the lowest order
(developed in Sec. 4) is not sufficient to even qualitatively reproduce
the meson-baryon interaction in the strange sector for L=1 partial wave.
The possible extension of the model that would be based on the $O(p^2)$ chiral Lagrangian is, in principle, possible,
but problematic due to the huge amount of new free parameters to consider.

\vspace{0.2in}
\emph{Acknowledgments}: I would like to thank P. Bedaque, A. Ciepl\'y, T.D. Cohen, and A. Gal for valuable discussions and encouragement.
This work was supported by the U.S.~Department of Energy through grant DE-FG02-93ER-40762.

\section*{Appendix}

The appendix consists of Tables \ref{tabC1}, \ref{tabC2}, and \ref{tabC3}, which specify the couplings $\mathcal{C}^{\dots}_{(ai)\rightarrow(bj)}$
among channels.

\begin{table}[h!]

\caption{Coupling matrix $\mathcal{C}^{WT}_{(ai)\rightarrow(bj)}$.}
\label{tabC1}

\begin{tabular}{|c|cccccccccc|}
\hline
	& $\pi^0 \Lambda$ & $\pi^0 \Sigma^0$ & $\pi^- \Sigma^+$ & $\pi^+ \Sigma^-$ & $K^- p$  & $\overline{K}^0 n$ & $\eta \Lambda$ & $\eta \Sigma^0$ & $K^0 \Xi^0$  & $K^+\Xi^-$ \\\hline
$\pi^0 \Lambda$ 	&  $0$ & $0$& $0$& $0$& $\sqrt{3}$& $-\sqrt{3}$& $0$& $0$& $-\sqrt{3}$& $\sqrt{3}$ \\
$\pi^0 \Sigma^0$ 	&       & $0$& $4$& $4$& $1$& $1$& $0$& $0$& $1$& $1$ \\
$\pi^- \Sigma^+$ 	&       &     & $4$& $0$& $2$& $0$& $0$& $0$& $2$& $0$ \\
$\pi^+ \Sigma^-$ 	&       &     &     & $4$& $0$& $2$& $0$& $0$& $0$& $2$ \\
$K^- p$  		&       &     &     &     & $4$& $2$& $3$& $\sqrt{3}$& $0$& $0$ \\
$\overline{K}^0 n$ 	&       &     &     &     &     & $4$& $3$& $-\sqrt{3}$& $0$& $0$ \\
$\eta \Lambda$  	&       &     &     &     &     &     & $0$& $0$& $3$& $3$ \\
$\eta \Sigma^0$ 	&       &     &     &     &     &     &     & $0$& $-\sqrt{3}$& $\sqrt{3}$ \\	
$K^0 \Xi^0$ 	&       &     &     &     &     &     &     &     & $4$& $2$ \\
$K^+\Xi^-$	&       &     &     &     &     &     &     &     &     & $4$ \\
\hline
\end{tabular}
\end{table}

\begin{sidewaystable}
\centering
\caption{Coupling matrix $\mathcal{C}^{direct}_{(ai)\rightarrow(bj)}$.}
\label{tabC2}

\begin{tabular}{|c|c|c|c|c|c|}
\hline
	& $\pi^0 \Lambda$ & $\pi^0 \Sigma^0$ & $\pi^- \Sigma^+$ & $\pi^+ \Sigma^-$ & $K^- p$   \\\hline
$\pi^0 \Lambda$ 	&  $4D^2$ & $0$& $-4\sqrt{3}DF$& $4\sqrt{3}DF$& $2\sqrt{3}D(D-F)$ \\
$\pi^0 \Sigma^0$ 	&       & $4D^2$& $4D^2$& $4D^2$& $-2D(D+3F)$ \\
$\pi^- \Sigma^+$ 	&       &     & $4D^2+12F^2$& $4D^2-12F^2$& $-2D(D+3F)-6F(D-F)$ \\
$\pi^+ \Sigma^-$ 	&       &     &     & $4D^2+12F^2$& $-2D(D+3F)+6F(D-F)$ \\
$K^- p$  		&       &     &     &     & $(D+3F)^2+3(D-F)^2$ \\\hline
&  $\overline{K}^0 n$ & $\eta \Lambda$ & $\eta \Sigma^0$ & $K^0 \Xi^0$  & $K^+\Xi^-$ \\\hline
$\pi^0 \Lambda$ 	&   $-2\sqrt{3}D(D-F)$& $0$& $4D^2$& $-2\sqrt{3}D(D+F)$& $2\sqrt{3}D(D+F)$ \\
$\pi^0 \Sigma^0$ 	&    $-2D(D+3F)$& $-4D^2$& $0$& $-2D(D-3F)$& $-2D(D-3F)$ \\
$\pi^- \Sigma^+$ 	&     $-2D(D+3F)+6F(D-F)$& $-4D^2$& $-4\sqrt{3}DF$& $-2D(D-3F)+6F(D+F)$& $-2D(D-3F)-6F(D+F)$ \\
$\pi^+ \Sigma^-$ 	&  $-2D(D+3F)-6F(D-F)$& $-4D^2$& $4\sqrt{3}DF$& $-2D(D-3F)-6F(D+F)$& $-2D(D-3F)+6F(D+F)$ \\
$K^- p$  		&   $(D+3F)^2-3(D-F)^2$& $2D(D+3F)$& $2\sqrt{3}D(D-F)$& $D^2-9F^2-3(D^2-F^2)$& $D^2-9F^2+3(D^2-F^2)$ \\
$\overline{K}^0 n$ 	&$(D+3F)^2+3(D-F)^2$& $2D(D+3F)$& $-2\sqrt{3}D(D-F)$& $D^2-9F^2+3(D^2-F^2)$& $D^2-9F^2-3(D^2-F^2)$ \\
$\eta \Lambda$  	&     & $4D^2$& $0$& $2D(D-3F)$& $2D(D-3F)$ \\
$\eta \Sigma^0$ 	&      &     & $4D^2$& $-2\sqrt{3}D(D+F)$& $2\sqrt{3}D(D+F)$ \\	
$K^0 \Xi^0$ 		&    &     &     & $(D-3F)^2+3(D+F)^2$& $(D-3F)^2-3(D+F)^2$ \\
$K^+\Xi^-$		     &     &     &     &     & $(D-3F)^2+3(D+F)^2$ \\\hline
\end{tabular}

\end{sidewaystable}

\begin{sidewaystable}
\centering
\caption{Coupling matrix $\mathcal{C}^{crossed}_{(ai)\rightarrow(bj)}$.}
\label{tabC3}

\begin{tabular}{|c|c|c|c|c|c|}
\hline
	& $\pi^0 \Lambda$ & $\pi^0 \Sigma^0$ & $\pi^- \Sigma^+$ & $\pi^+ \Sigma^-$ & $K^- p$   \\\hline
$\pi^0 \Lambda$ 	&  $4D^2$ & $0$& $4\sqrt{3}DF$& $-4\sqrt{3}DF$& $-\sqrt{3}(D+F)(D+3F)$ \\
$\pi^0 \Sigma^0$ 	&       & $4D^2$& $-12F^2$& $-12F^2$& $3(D+F)(D-F)$\\
$\pi^- \Sigma^+$ 	&       &     & $0$& $4D^2-12F^2$& $0$ \\
$\pi^+ \Sigma^-$ 	&       &     &     & $0$& $6(D+F)(D-F)$ \\
$K^- p$  		&       &     &     &     & $0$ \\\hline
	&  $\overline{K}^0 n$ & $\eta \Lambda$ & $\eta \Sigma^0$ & $K^0 \Xi^0$  & $K^+\Xi^-$ \\\hline
$\pi^0 \Lambda$ 	& $\sqrt{3}(D+F)(D+3F)$& $0$& $-4D^2$& $\sqrt{3}(D-F)(D-3F)$& $-\sqrt{3}(D-F)(D-3F)$ \\
$\pi^0 \Sigma^0$ 	& $3(D+F)(D-F)$& $4D^2$& $0$& $3(D-F)(D+F)$& $3(D-F)(D+F)$ \\
$\pi^- \Sigma^+$ 	&   $6(D+F)(D-F)$& $4D^2$& $-4\sqrt{3}DF$& $0$& $6(D-F)(D+F)$ \\
$\pi^+ \Sigma^-$ 	& $0$& $4D^2$& $4\sqrt{3}DF$& $6(D-F)(D+F)$& $0$ \\
$K^- p$  		&  $0$& $(D+3F)(D-3F)$& $-\sqrt{3}(D-F)(D-3F)$& $6(D+F)(D-F)$& $D^2-9F^2+3(D^2-F^2)$ \\
$\overline{K}^0 n$ 	     & $0$& $(D+3F)(D-3F)$& $\sqrt{3}(D-F)(D-3F)$& $D^2-9F^2+3(D^2-F^2)$& $6(D+F)(D-F)$ \\
$\eta \Lambda$  	     &     & $4D^2$& $0$& $(D+3F)(D-3F)$& $(D+3F)(D-3F)$ \\
$\eta \Sigma^0$ 	     &     &     & $4D^2$& $\sqrt{3}(D+F)(D+3F)$& $-\sqrt{3}(D+F)(D+3F)$ \\	
$K^0 \Xi^0$ 		   &     &     &     & $0$& $0$ \\
$K^+\Xi^-$	   &     &     &     &     & $0$ \\\hline

\end{tabular}

\end{sidewaystable}


\newpage 


\begin{thebibliography}{99}

\bibitem{Weinberg} S. Weinberg, Physica {\bf 96A} (1979) 327.
\bibitem{BernardKaiserMeissner} V. Bernard, N. Kaiser, U.G. Mei\ss ner, Int. J. Mod. Phys. E {\bf 4} (1995) 193.
\bibitem{PDG} K. Nakamura and Particle Data Group 2010, J. Phys. {\bf G 37} (2010) 075021.

\bibitem{KaiserSiegelWeise} N. Kaiser, P.B. Siegel, W. Weise, Nucl. Phys. A {\bf 594} (1995) 325.
\bibitem{CieplySmejkal07} A. Ciepl\'y, J. Smejkal, Eur. Phys. J. A {\bf 34} (2007) 237.
\bibitem{CieplySmejkal10} A. Ciepl\'y, J. Smejkal, Eur. Phys. J. A {\bf 43} (2010) 191.


\bibitem{WaasKaiserWeise} T. Waas, N. Kaiser, W. Weise, Phys. Lett. B {\bf 365} (1996) 12.
\bibitem{Lutz} M. Lutz, Phys. Lett. B {\bf 426} (1998) 12.

\bibitem{CieplyFriedman..} A. Ciepl\'y, E. Friedman, A. Gal, D. Gazda, J. Mare\v{s}, Phys. Rev. C {\bf 84} (2011) 045206.
\bibitem{CieplySmejkal12} A. Ciepl\'y, J. Smejkal,	Nucl. Phys. A {\bf 881} (2012) 115.

\bibitem{KrejcirikCieplyGal} V. Krej\v{c}i\v{r}\'ik, A. Ciepl\'y, A. Gal, Phys. Rev. C {\bf 82} (2010) 024609.
\bibitem{CieplyFriedmanGalKrejcirik} A. Ciepl\'y, E. Friedman, A. Gal, V. Krej\v{c}i\v{r}\'ik, Phys. Lett. B {\bf 698} (2011) 226.

\bibitem{WeiseHartle} W. Weise, R. H\"artle, Nucl. Phys. A {\bf 804} (2008) 173.
\bibitem{GazdaMares} D. Gazda, J. Mare\v{s}, arXiv:1206.0223v1 [nucl-th].

\bibitem{Goldberger} M.L. Goldberger, K.M. Watson, {\it Collision theory}, John Willey \& Sons (1964).

\bibitem{OllerMeissner} J.A. Oller, U.G. Mei\ss ner, Phys. Lett. B {\bf 500} (2001) 263.
\bibitem{BorasoyNisslerWeise} B. Borasoy, R. Ni\ss ler, W. Weise,  Eur. Phys. J. A {\bf 25} (2005) 79.
\bibitem{BrunsMaiMeissner} P.C. Bruns, M. Mai,  U.G. Mei\ss ner, Phys. Lett. B {\bf 697} (2011) 254.
\bibitem{MaiMeissner} M. Mai,  U.G. Mei\ss ner, arXiv:1202.2030v1 [nucl-th].

\bibitem{Yamaguchi1} Y. Yamaguchi, Phys. Rev. {\bf 95} (1954) 1628.

\bibitem{ADMartin} A.D. Martin, Nucl. Phys. B {\bf 179} (1981) 33.
\bibitem{BazziEtAl} M. Bazzi et al., Nucl. Phys. A {\bf 881} (2012) 88.

\bibitem{Ciborowski} J. Ciborowski et al., J. Phys. G {\bf 8} (1982) 13.
\bibitem{Evans} D. Evans et al., J. Phys. G {\bf 9} (1983) 885.
\bibitem{Sakitt} M. Sakitt et al., Phys. Rev. {\bf 139} (1965) 719.
\bibitem{HumphreyRoss} W.E. Humphrey, R.R. Ross, Phys. Rev. {\bf 127} (1962) 1305.
\bibitem{Mast} T.S. Mast et al., Phys. Rev. D {\bf 14} (1976) 13.
\bibitem{Bangerter} R.O. Bangerter et al., Phys. rev. D {\bf 23} (1981) 1485.
\bibitem{Watson} M.B. Watson, M. Ferro-Luzzi, R.D. Tripp, Phys. Rev. {\bf 131} (1963) 2248.
\bibitem{Watson} M.B. Watson, M. Ferro-Luzzi, R.D. Tripp, Phys. Rev. {\bf 131} (1963) 2248.
\bibitem{JKKim} J.K. Kim, Phys. Rev. Lett. {\bf 19} (1967) 1074.

\end{thebibliography}
\end{document}